\documentclass{cernrep}

\def\GeV{{\rm GeV}}

\def\ttt#1{\texttt{\small #1}}

\providecommand{\gaga}{\gamma\,\gamma}

\newcommand{\bbbar}{\ensuremath{b\bar{b}}}
\newcommand{\sqrts}{\sqrt{s}}

\newcommand{\sqrtsnn}{\sqrt{s_{_{\textsc{nn}}}}}

\newcommand{\bbar}     {\ensuremath{b\bar{b}}}

\newcommand{\qqbar}    {\ensuremath{q\bar{q}}}
\newcommand{\ccbar}    {\ensuremath{c\bar{c}}}
\providecommand{\madgraph}{{\sc madgraph}}
\providecommand{\pythia}{{\sc pythia}}

\providecommand{\hdecay}{{\sc hdecay}}
\providecommand{\fastjet}{{\sc fastjet}}
\newcommand{\LumiInt}{\mathcal{L}_{\rm \tiny{int}}}

\newcommand*{\cm}{c.m.\@\xspace}

\begin{document}
\title{Prospects for $\gamma\gamma\to$~Higgs observation in ultraperipheral ion collisions at the Future Circular Collider}
\author{David d'Enterria$^{1}$, Daniel E. Martins$^{2}$, and Patricia Rebello Teles$^{3}$}
\institute{$^{1}$CERN, EP Department, 1211 Geneva. \\ 
$^{2}$UFRJ, Univ. Federal do Rio de Janeiro, 21941-901, Rio de Janeiro, RJ. \\ 
$^{3}$CBPF, Centro Brasileiro de Pesquisas F\'{i}sicas, 22290-180, Rio de Janeiro, RJ.}

\begin{abstract}
We study the two-photon production of the Higgs boson, $\rm \gaga\to H$, 
at the Future Circular Collider (FCC) in ultraperipheral PbPb and pPb collisions at 
$\sqrtsnn = 39$ and 63~TeV. Signal and background events are generated with 
\madgraph~5, including $\gamma$ fluxes from the proton and lead ions in the equivalent 
photon approximation, yielding $\rm \sigma(\gaga\to H)$~=~1.75~nb and 1.5~pb in
PbPb and pPb collisions respectively. We analyse the H$\to\bbar$ decay mode
including realistic reconstruction efficiencies for the final-state $b$-jets, 
showered and hadronized with \pythia~8, as well as appropriate selection criteria 
to reduce the $\gaga\to\bbar,\ccbar$ continuum backgrounds.
Observation of PbPb$\xrightarrow{\gamma\gamma}$(Pb)H(Pb) 
is achievable in the first year with the expected FCC integrated luminosities.
\end{abstract}


\maketitle

\section{Introduction}

The observation of the predicted Higgs boson~\cite{Higgs} in proton-proton collisions 
at the Large Hadron Collider~\cite{Chatrchyan:2012xdj,Aad:2012tfa} 
has represented a breakthrough in our scientific understanding of the particles and forces in nature.
A complete study of the properties of the scalar boson, including its couplings to all known particles, and searches of possible 
deviations indicative of physics beyond the Standard Model (SM), require a new collider facility with much higher 
center-of-mass (\cm) energies~\cite{dEnterria:2017dac}. 
The Future Circular Collider (FCC) is a post-LHC project at CERN, aiming at pp collisions 
up to at a \cm\ energy of $\sqrt{s} = 100$~TeV in a new 80--100~km tunnel with 16--20~T dipoles~\cite{Mangano:2016jyj}. 
The FCC running plans with hadron beams (FCC-hh) includes also heavy-ion operation at nucleon-nucleon \cm\ energies 
of $\sqrt{100}~{\rm TeV}. \sqrt{Z_{1}Z_{2}/(A_{1}A_{2})}=39~{\rm TeV}, 63~{\rm TeV}$ for PbPb, pPb with 
(monthly) integrated luminosities of 110~nb$^{-1}$ and 29~pb$^{-1}$~\cite{Dainese:2016gch}. Such high collision
energies and luminosities, factors of 7 and 30 times higher respectively than those reachable at the LHC, open
up the possibility to study the production of the Higgs boson in nuclear collisions, both in central 
hadronic~\cite{dEnterria:2017jyt} as well as in ultraperipheral (electromagnetic)~\cite{dEnterria:2009cwl} 
interactions. The observation of the latter $\gaga\to$~H process provides an independent measurement of 
the H-$\gamma$ coupling not based on Higgs decays but on its $s$-channel production mode.

\begin{figure}[htbp!]
\begin{center}
\includegraphics[width=8.2cm,height=6.35cm]{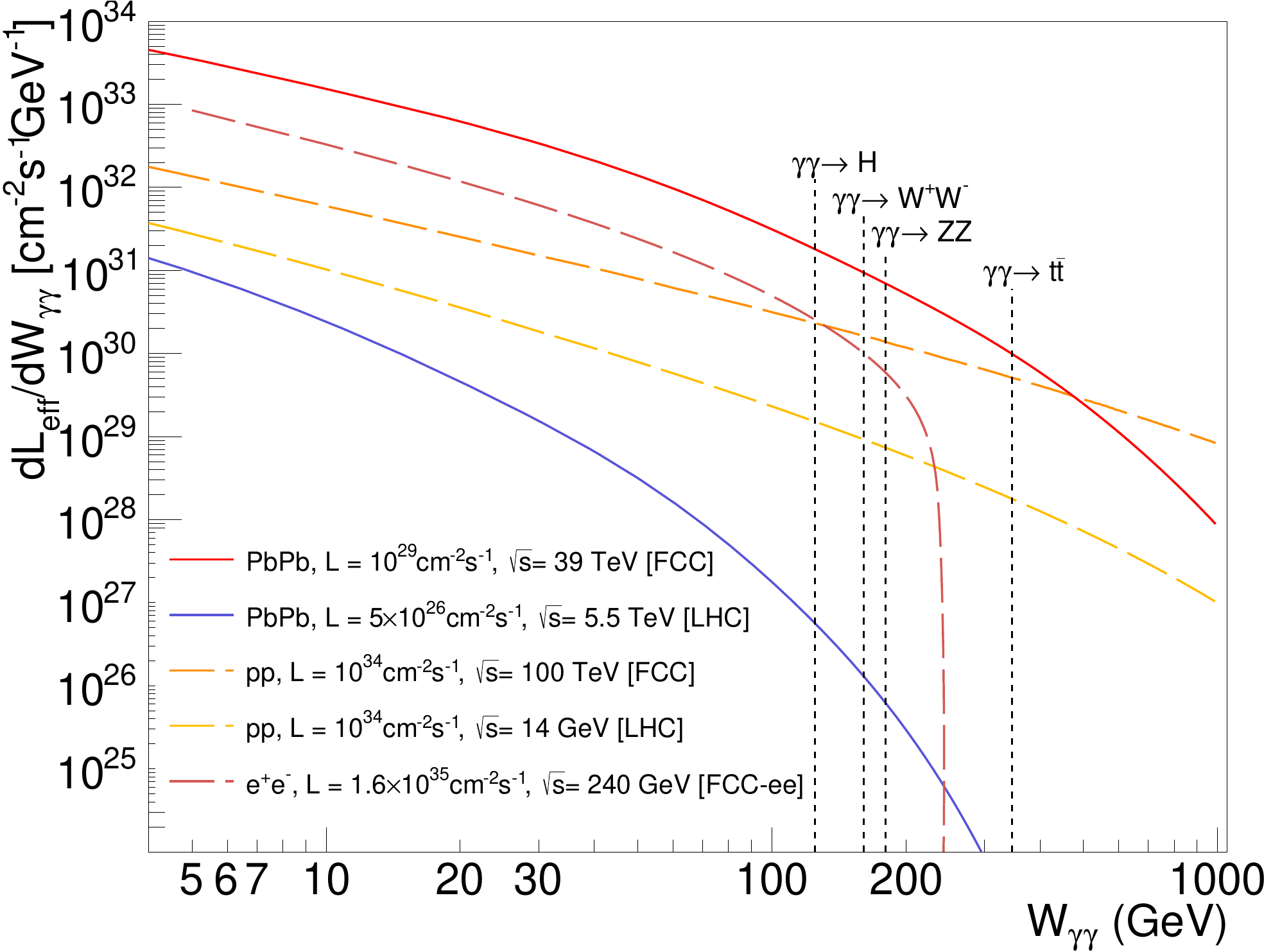}
\includegraphics[width=7.7cm]{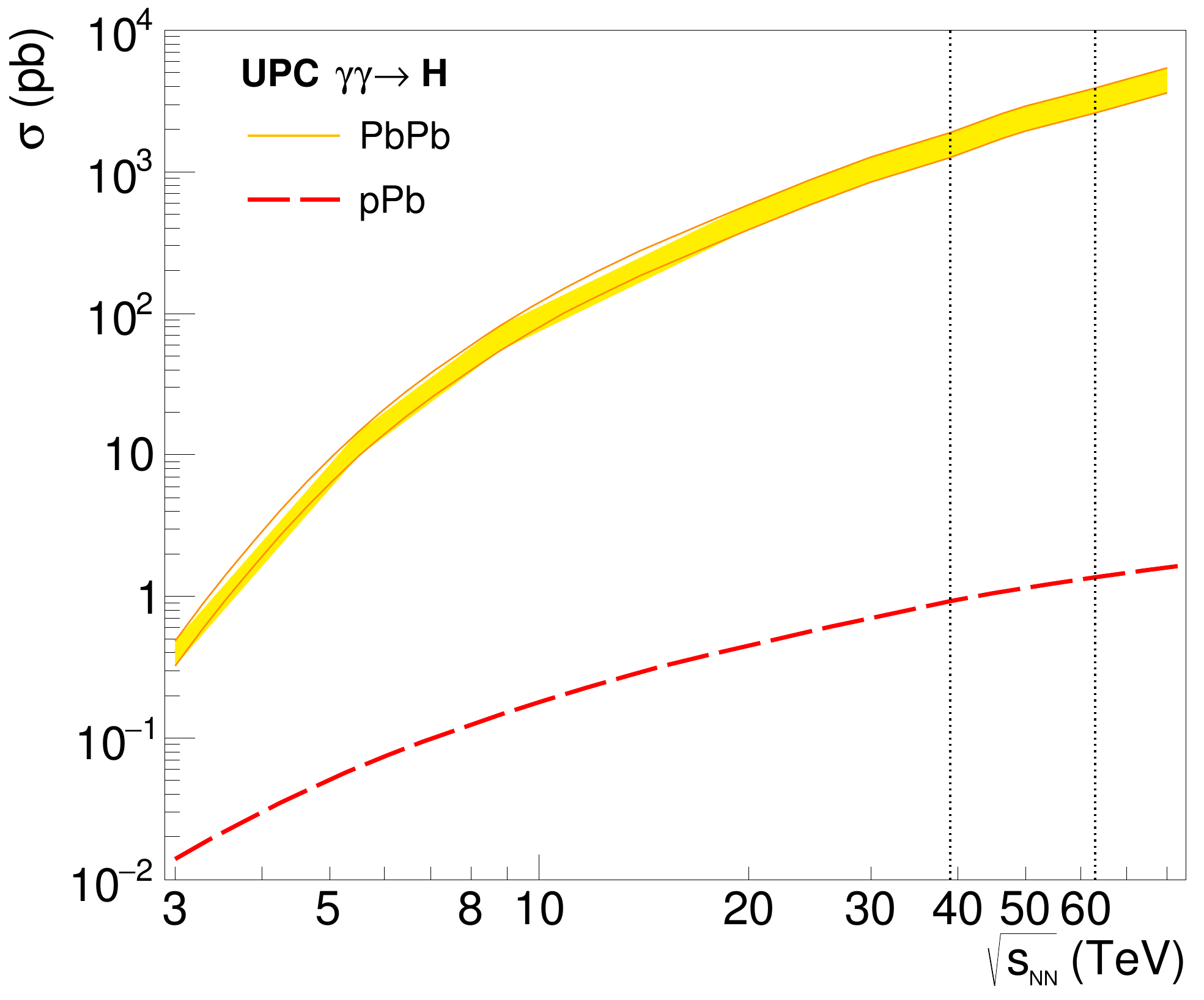}
\caption{ Left: Two-photon effective luminosities  as a function of $\gaga$ \cm\ energy over 
$W_{\gaga} \approx$~5--1000~GeV in PbPb, pp, and $e^+e^-$ collisions at the FCC~\cite{Mangano:2016jyj,Dainese:2016gch,eds17}, 
and in PbPb and pp collisions at the LHC.
Right: Two-photon fusion Higgs boson cross section versus nucleon-nucleon \cm\ energy in ultraperipheral PbPb (top) 
and pPb (bottom curve) collisions. The vertical lines indicate the expected FCC running energies 
at $\sqrtsnn$~=~39 and 63~TeV. }
\label{fig:1}
\end{center}
\end{figure}

The measurement of exclusive $\rm \gaga\to H$ in ultraperipheral collisions (UPCs)~\cite{UPCs} 
of pPb and PbPb beams has been studied in detail for LHC energies\footnote{A few older papers had already previously discussed the
possibility to produce the Higgs boson in heavy-ion UPCs~\cite{higgs_upc}.}  
in ref.~\cite{dEnterria:2009cwl}, although its observation there is unfeasible with the nominal luminosities 
(Fig.~\ref{fig:1}, left). We extend  
such studies for FCC energies, where such an observation is warranted. All charges accelerated at high energies generate 
electromagnetic fields which, in the equivalent photon approximation (EPA)~\cite{WW}, can be considered as quasireal
photon beams\footnote{The emitted photons are almost on mass shell, with virtuality $- Q^{2} < 1/R^{2}$, 
where $R$ is the radius of the charge, i.e. $Q\approx$~0.28~GeV for protons ($R\approx$~0.7~fm) and $Q<$~0.06~GeV 
for nuclei ($R_A\approx 1.2\,A^{1/3}$~fm) with mass number $A>$~16.}~\cite{Stan70}. 
The highest available photon energies are of the order of the inverse Lorentz-contracted radius $R$ of the source charge, 
$\omega_{\rm max}\approx\gamma/R$, which at the FCC yield photon-photon collisions above 1~TeV (Table~\ref{tab:1}).
In addition, since the photon flux scales as the squared charge of the beam, $Z^2$, two-photon cross sections are 
enhanced millions of times for ions ($Z_{\rm Pb}^4$~=~5$\cdot$10$^{7}$ for PbPb) compared to proton or electron 
beams, thereby featuring the largest $\gaga$ luminosities among all colliding systems (Fig.~\ref{fig:1}, left).
\begin{table}[htpb]
\begin{center}
\caption[]{Relevant parameters for photon-photon processes in ultraperipheral pPb and PbPb collisions at the FCC:
(i) nucleon-nucleon \cm\ energy, $\sqrtsnn$, (ii) integrated luminosity per year, $\LumiInt$, 
(iii) beam energies, $\rm E_{beam}$, (iv) Lorentz factor, $\gamma_{_{\rm L}}=\sqrtsnn/(2\,m_N)$,  
(v) effective (Pb) radius, $R_A$, 
(vi) photon ``cutoff energy'' in the \cm\ frame, $\omega_{\rm max}$, and
(vii) ``maximum'' photon-photon \cm\ energy, $\sqrt{s_{\gaga}^{\rm max}}$.
The last column lists the $\rm \gaga\to H$ cross sections.}
\label{tab:1}
\vspace{-0.25cm}
\begin{tabular}{lcccccccc} \hline
System\!\!  & \!\!$\sqrtsnn$\!\! & \!\!$\LumiInt$\!\! & \!\!\!\!$E_{\rm beam1}+E_{\rm beam2}$\!\! & $\gamma_{_{\rm L}}$ &  $R_A$ & $\omega_{\rm max}$ 
       & $\sqrt{s_{\gaga}^{\rm max}}$ & $\rm \sigma(\gamma\gamma\to H)$ \\
\hline
pPb & 63 TeV  &   29~pb$^{-1}$ & 50. + 19.5 TeV  & 33\,580 & 7.1 fm & 950 GeV & 1.9 TeV &  1.5~pb\\ 
PbPb & 39 TeV &  110~nb$^{-1}$ & 19.5 + 19.5 TeV & 20\,790 & 7.1 fm & 600 GeV & 1.2 TeV & 1.75~nb\\ \hline
\end{tabular}
\end{center}
\end{table}

\vspace{-0.5cm}
\section{Theoretical setup}
\label{sec:TH}

The \madgraph~5 (v.2.5.4)~\cite{madgraph} Monte Carlo (MC) event generator is used to compute the relevant 
cross sections from the convolution of the Weizs\"acker-Williams EPA photon fluxes~\cite{WW} for the proton and lead ion, 
and the H-$\gamma$ coupling parametrized in the Higgs effective field theory~\cite{heft},
following the implementation discussed in~\cite{dEnterria:2009cwl} with a more accurate treatment of the 
non hadronic-overlap correction. 
The proton $\gamma$ flux is given by the energy spectrum $f_{\gamma/p}(x)$ where $x = \omega/E$ 
is the fraction of the beam energy carried by the photon~\cite{Budnev:1974de}:
\begin{equation}
f_{\gamma/p}(x) = \frac{\alpha}{\pi} \, \frac{1 - x + 1/2 x^2}{x} 
\int_{Q_{\rm min}^2}^{\infty} \frac{Q^2 - Q_{\rm min}^2}{Q^4} | F(Q^2) |^2 dQ^2 \;,
\label{eq:f_x}
\end{equation}
with $\alpha=1/137$, $F(Q^2)$ the proton electromagnetic form factor, and the minimum momentum transfer $Q_{\rm min}$
is a function of $x$ and the proton mass $m_p$, $Q_{\rm min}^2 \approx (x m_p)^2/(1-x)$.
The photon energy spectrum of the lead ion ($Z=82$), integrated over impact parameter $b$ from $b_{\rm min}$
 to infinity, is given by~\cite{Jackson}:
\begin{equation}
f_{\gamma/Pb}(x) = \frac{\alpha Z^2}{\pi} \, \frac{1}{x} \, \bigg[ 2 x_i K_0(x_i) K_1(x_i) - x_i^2 (K_1^2(x_i) - K_0^2(x_i)) \bigg] \; ,
\label{eq:flux_A}
\end{equation}
where $x_i= x\, m_N \, b_{\rm min}$, and $K_0$, $K_1$ are the modified Bessel functions of the second kind
of zero and first order, related respectively to the emission of longitudinally and transversely polarized photons. 
The latter dominating for ultrarelativistic charges ($\gamma_{_{\rm L}}\gg1$). 
The dominant Higgs decay mode is $\rm H\to\bbar$, with a branching fraction of 58\% as computed with \hdecay~\cite{hdecay}.
The \pythia 8.2~\cite{pythia8} MC generator was employed to shower and hadronize the two final-state $b$-jets,
which are then reconstructed with the Durham $k_{t}$ algorithm~\cite{kTalgo} (exclusive 2-jets final-state) using 
\fastjet~3.0~\cite{fastjet}.
The same setup is used to generate the exclusive two-photon production of $\bbar$ and (possibly misidentified) 
$\ccbar$ and light-quark ($\qqbar$) jet pairs, which constitute the most important physical background for the measurement 
of the H$\to \bbar$ channel.

\section{Results}

The total elastic Higgs boson cross sections in ultraperipheral PbPb and pPb collisions as a function of $\sqrts$ are
shown in Fig.~\ref{fig:1} (right). We have assigned a conservative 20\% uncertainty to the predicted cross sections to cover
different charge form factors. At LHC energies, we find a slightly reduced cross section, 
$\rm \sigma(PbPb\to\gaga\to H) = 15 \pm 3$~~pb, compared to the results of~\cite{dEnterria:2009cwl} 
due a more accurate treatment of the non hadronic-overlap correction based on~\cite{Klein:2016yzr}.
The predicted total Higgs boson cross sections are $\rm \sigma(\gaga\to H)$~=~1.75~nb and 1.5~pb in
PbPb and pPb collisions at $\sqrtsnn$~=~39 and 63~TeV which, for the nominal $\LumiInt$~=~110~nb$^{-1}$
and 29~pb$^{-1}$ luminosities per ``year'' (1-month run), imply $\sim$200 and 45 Higgs bosons 
produced (corresponding to 110 and 25 bosons in the $\bbar$ decay mode, respectively). The main backgrounds 
are pairs from the $\gamma\gamma \rightarrow \bbar, \ccbar, \qqbar$ continuum, where charm and light 
($q=uds$) quarks are misidentified as $b$-quarks. 
The irreducible $\gaga \to b\bar{b}$ background over the mass range $100 < W_{\gaga} < 150$~GeV
is $\sim$20 times larger than the signal, but can be suppressed (as well as that from misidentified $c\bar{c}$ 
and $q\bar{q}$ pairs) via various kinematical cuts. The data analysis follows closely the similar 
LHC study~\cite{dEnterria:2009cwl}, with the following reconstruction performances assumed: jet reconstruction 
over $|\eta|<5$, 7\% $b$-jet energy resolution (resulting in a dijet mass resolution of $\sim$6~GeV at the Higgs peak), 
70\% $b$-jet tagging efficiency, and 5\% (1.5\%) $b$-jet mistagging probability for a $c$ (light-flavour $q$) 
quark. For the double $b$-jet final-state of interest, these lead to a $\sim$50\% efficiency for the MC-generated signal (S), 
and a total reduction of the misidentified $c\bar{c}$ and $q\bar{q}$ continuum backgrounds (B) by factors of 
$\sim$400 and $\sim$400\,000.

\begin{table}[htbp!]
\begin{center}
\caption{Summary of the cross sections and expected number of events per run after event selection and reconstruction
criteria (see text) for signal and backgrounds in the ${\rm \gaga \rightarrow H}(\bbar)$ analysis, 
obtained from events generated with \madgraph~5+\pythia~8 for PbPb and pPb collisions at FCC energies.}
\label{tab:2}
\begin{tabular}{lccc}\hline
PbPb at $\sqrtsnn$ = 39 TeV             & \!\!cross section                      & visible cross section & $N_{\rm evts}$  \\
             & \!\!\!\!($b$-jet (mis)tag efficiency) & after $p_T^j,\cos \theta_{jj},m_{jj}$ cuts & ($\LumiInt$~=~110 nb$^{-1}$) \\\hline
$\gaga \rightarrow {\rm H} \rightarrow \bbar$                               & 1.02~nb (0.50~nb)& 0.19 nb & 21.1 \\
$\gaga \rightarrow \bbar$  \;$[\rm \scriptstyle{m_{\bbar} = 100-150~GeV]}$  & 24.3~nb (11.9~nb)& 0.23 nb & 25.7 \\
$\gaga \rightarrow \ccbar$ \;$[\rm \scriptstyle{m_{\ccbar} = 100-150~GeV]}$ & 525~nb (1.31~nb) & 0.02 nb &  2.3 \\
$\gaga \rightarrow \qqbar$ \;$[\rm \scriptstyle{m_{\qqbar} = 100-150~GeV]}$ & 590~nb (0.13~nb) & 0.002 nb& 0.25 \\\hline
pPb at $\sqrtsnn$ = 63 TeV              &  &  & $N_{\rm evts}$ \\
             & & & ($\LumiInt$~=~29 pb$^{-1}$) \\\hline
$\gaga \rightarrow {\rm H} \rightarrow \bbar$                              & 0.87~pb (0.42~pb) & 0.16 pb & 4.8 \\
$\gaga \rightarrow \bbar$  \;$[\rm \scriptstyle{m_{\ensuremath\bbar} = 100-150~GeV]}$ & 21.8~pb (10.7~pb) & 0.22 pb & 6.3 \\
$\gaga \rightarrow \ccbar$ \;$[\rm \scriptstyle{m_{\ccbar} = 100-150~GeV]}$ & 410.~pb (1.03~pb) & 0.011 pb& 0.3 \\
$\gaga \rightarrow \qqbar$ \;$[\rm \scriptstyle{m_{\qqbar} = 100-150~GeV]}$ & 510.~pb (0.114~pb)& 0.001 pb& 0.04 \\\hline
\end{tabular}
\end{center}
\end{table}
As proposed in~\cite{dEnterria:2009cwl}, various simple kinematical cuts can be applied to enhance the S/B ratio.
Since the transverse momenta of the Higgs decay $b$-jets peak at $p_{T}^{j} \approx m_{\rm H}/2$, selecting events 
with at least one jet within $p_{T}$~=~55--62.5~GeV suppresses $\sim$96\% of the continuum backgrounds, 
while removing only half of the signal. Also, one can exploit the fact that the angular distribution of
the Higgs decay $b$-jets in the helicity frame is isotropically distributed in $|\cos \theta_{j_{1}j_{2}}|$, 
\ie\ each jet is independently emitted either in the same direction as the $\bbar$ pair or opposite to it, 
while the continua (with quarks propagating in the $t$- or $u$- channels) are peaked in the forward--backward 
directions. Thus, requiring $|\cos \theta_{j_{1}j_{2}}| < 0.5$ further suppresses the continuum contaminations 
by another $\sim$20\% while leaving untouched the  signal. The significance of the signal can then be computed 
from the final number of counts within 1.4$\sigma$ around the Gaussian Higgs peak (\ie\ $117 < m_{\bbbar} < 133$~GeV) 
over the underlying dijet continuum. Table~\ref{tab:2} summarizes the visible cross sections and the number of events 
after cuts for the nominal luminosities of each system.

\begin{figure}[h]
\centering
\includegraphics[width=7.95cm]{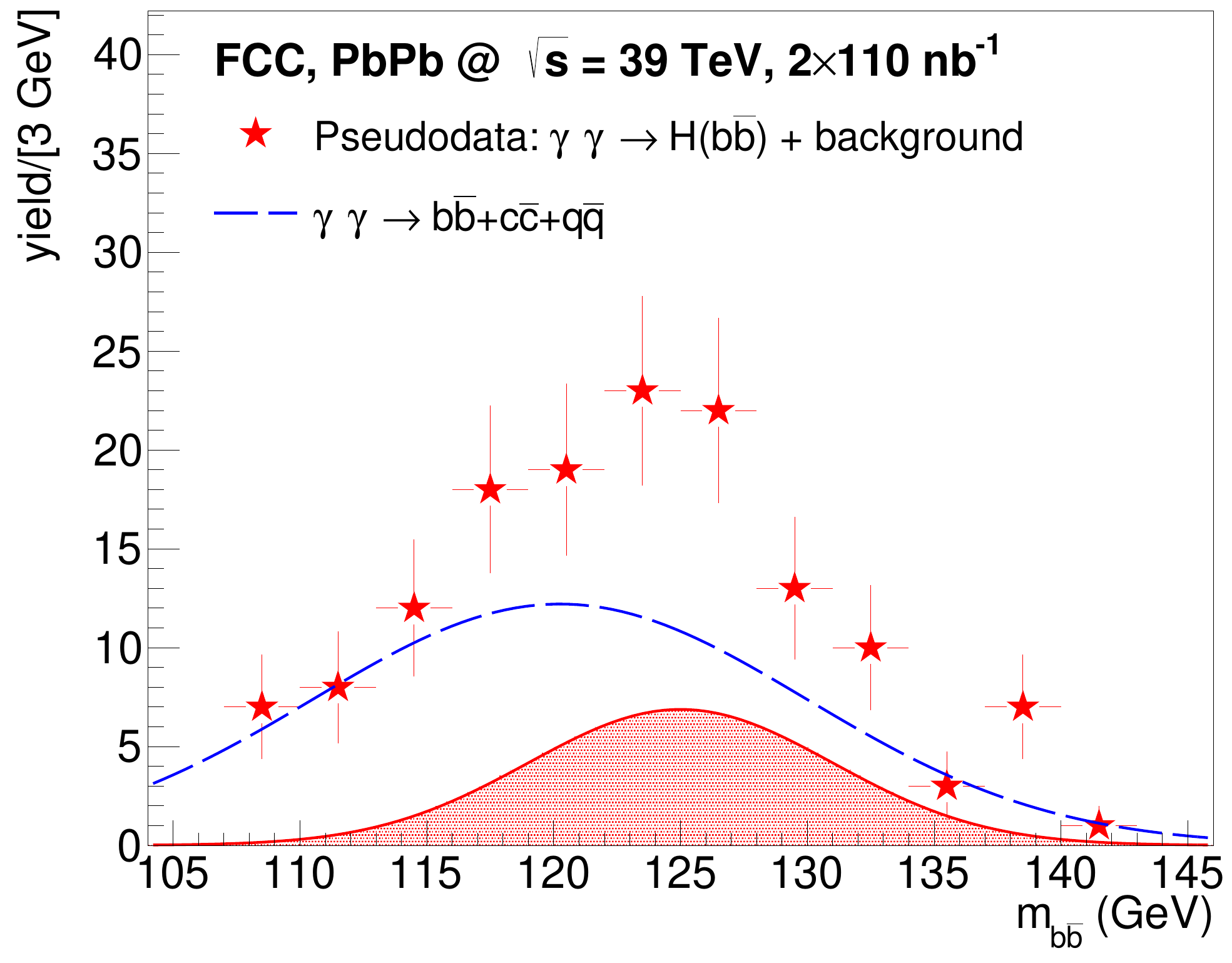}
\includegraphics[width=7.95cm]{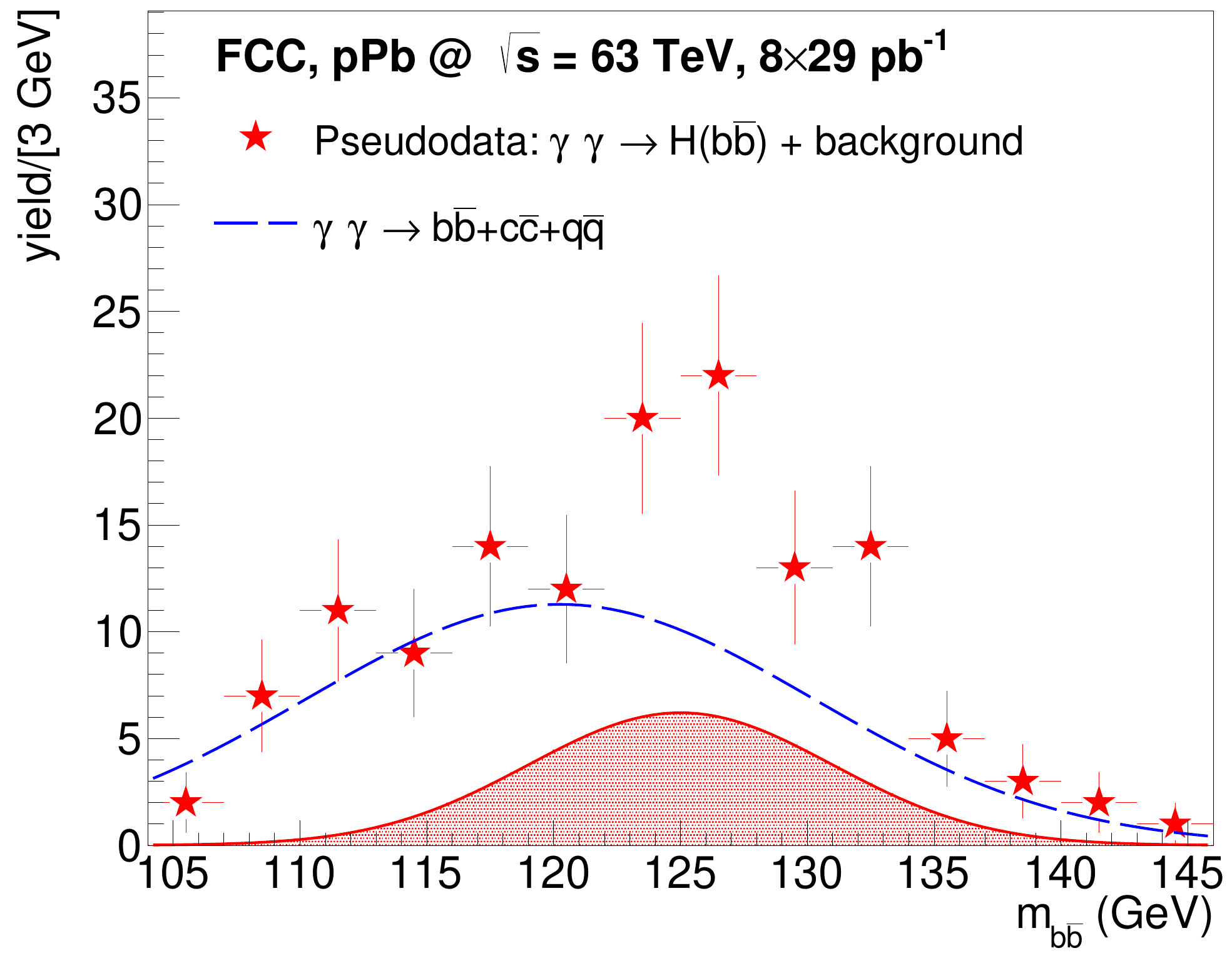}
\caption{\label{fig:2}Expected dijet invariant mass distributions for photon-fusion
H($\bbar$) signal (hatched Gaussian) and $\bbar+\ccbar+\qqbar$ continuum (dashed line) in ultraperipheral 
PbPb ($\sqrtsnn = 39$~TeV, left) and pPb ($\sqrtsnn = 63$~TeV, right) collisions, after event selection 
and reconstruction criteria  (see text) with the quoted integrated luminosities.}
\end{figure}

In PbPb $\sqrts$~=~39 GeV for the nominal integrated luminosity of $\LumiInt = 110\;\mbox{nb}^{-1}$
per run, we expect about $\sim$21 signal counts over $\sim$28 for the sum of backgrounds in a window
$m_{\bbbar}$~=~117--133~GeV around the Higgs peak. Reaching a statistical significance close to 
5$\sigma$ (Fig.~\ref{fig:2}, left) would require to combine two different experiments 
(or doubling the luminosity in a single one).
Similar estimates for pPb at 63~TeV (29~pb$^{-1}$) yield about 5 signal events after cuts, on top of a background
of 6.7 continuum events. Reaching a 5$\sigma$ significance for the observation of $\gaga\to$~H production
(Fig.~\ref{fig:2}, right) would require in this case to run for about 8 months (instead of the nominal 
1-month run per year), or running 4 months and combining two experiments.
All the derived number of events and significances are based on the aforementioned set of kinematical cuts, 
and can be likely improved by using a more advanced multivariate analysis.

\section{Summary}

We have presented prospect studies for the measurement of the two-photon production of the Higgs boson in the $\bbar$ 
decay channel in ultraperipheral PbPb and pPb collisions at the FCC. Cross sections have been obtained 
at nucleon-nucleon \cm\ energies of $\sqrtsnn = 39$ and 63~TeV with \madgraph~5, using the Pb (and proton) 
equivalent photon fluxes and requiring no hadronic overlap of the colliding particles. 
The $b$-quarks have been showered and hadronized with
\pythia~8, and reconstructed in a exclusive two-jet final-state with the $k_T$ algorithm. By assuming 
realistic jet reconstruction performances
and (mis)tagging efficiencies, and applying appropriate kinematical cuts on the jet $p_T$ and dijet mass and
angles in the helicity frame, we can reconstruct the H$(\bbar)$ signal on top of the dominant $\gaga\to\bbar$
continuum background. The measurement of ${\rm \gaga \rightarrow H} \rightarrow \bbar$ would yield 21 (5) signal 
counts over 28 (7) continuum dijet pairs around the Higgs peak, in PbPb (pPb) collisions for their nominal integrated
luminosities per run. Observation of the photon-fusion Higgs production at the $5\sigma$-level is achievable in the
first year by combining the measurements of two experiments (or doubling the luminosity in a single one) in PbPb, 
and by running for about 8 months (or running 4 months and combining two experiments) in the pPb case.
The feasibility studies presented here confirm the interesting Higgs physics potential open to study 
in $\gamma\gamma$ ultraperipheral ion collisions at the FCC, providing an independent measurement of the H-$\gamma$ 
coupling not based on Higgs decays but on a $s$-channel production mode.\\

\noindent {\bf Acknowledgments --} P.\,R.\,T. acknowledges financial support from the CERN TH Department and from the FCC project.

\end{document}